# Thermo-mechanical properties of nitrogenated holey graphene (C$_2$N): A comparison of machine-learning-based and classical interatomic potentials

Saeed Arabha and Ali Rajabpour[*]

*Advanced Simulation and Computing Laboratory (ASCL), Mechanical Engineering Department, Imam Khomeini International University, Qazvin, Iran*

**Abstract**

Thermal and mechanical properties of two-dimensional nanomaterials are commonly studied by calculating force constants using the density functional theory (DFT) and classical molecular dynamics (MD) simulations. Although DFT simulations offer accurate estimations, the computational cost is high. On the other hand, MD simulations strongly depend on the accuracy of interatomic potentials. Here, we investigate thermal conductivity and elastic modulus of nitrogenated holey graphene (C$_2$N) using passively fitted machine-learning interatomic potentials (MLIPs), which depend on computationally inexpensive ab-initio molecular dynamics trajectories. Thermal conductivity of C$_2$N is investigated via MLIP-based non-equilibrium molecular dynamics simulations (NEMD). At room temperature, the lattice thermal conductivity of 85.5 ± 5 W/m-K and effective phonon mean free path of 37.16 ± 3 nm are found. By carrying out uniaxial tension simulations, the elastic modulus, ultimate strength, and fractural strain of C$_2$N are predicted to be 390 ± 3 GPa, 42 ± 2 GPa, and 0.29 ± 0.01, respectively. It is shown that the passively fitted MLIPs can be employed as an efficient interatomic potential to obtain the thermal conductivity and elastic modulus of C$_2$N utilizing classical MD simulations. Moreover, the possibility of employing MLIPs to simulate C$_2$N with point defects has been investigated. By training MLIP with point defect configurations, the mechanical properties of defective structures were studied. Although using the MLIP is more costly than classical interatomic potentials, it could efficiently predict the thermal and mechanical properties of 2D nanostructures.

---

[*] rajabpour@eng.ikiu.ac.ir

# 1. Introduction

Two-dimensional (2D) nanostructures have attracted significant attention in recent years[1]. Materials such as graphene[2], borophene[3], phosphorene[4], and many other structures are synthesized in single- and few-layer forms. These materials have a full range of electronic characteristics including metals, semiconductors, and insulators. Due to their sheet-like structure, most of the 2D nanomaterials are completely resistant to tensile strain while they have almost no resistance to compressive strain. Meanwhile, 2D nanostructures cover a wide range of heat transport properties, from low to very high values. Furthermore, the properties of these nanostructures, including mechanical and thermal properties, are well adjustable[5–7].

Recently, a promising 2D nanostructure known as $C_2N$ has been synthesized directly from graphene sheets[8]. The atomic structure of nitrogen-containing porous graphene consists of rings of carbon atoms terminated by nitrogen atoms. $C_2N$ porous monolayer is an excellent option for separating 'He' from natural gas due to its excellent permeability[9]. It can also be used as a promising thermoelectric material and provide theoretical guidance for fabricating outstanding thermoelectric devices[10,11].

Due to the atomic thicknesses of 2D nanomaterials, experimental measurement of mechanical and thermal properties of these nanostructures is quite challenging. Today, to study the properties of 2D nanostructures, accurate theoretical approaches are of great importance. In this regard, classical molecular dynamics (MD) simulations and density functional theory plus the Boltzmann equation of transport (DFT-BTE) are the most common methods for calculating the properties of these nanostructures, such as mechanical properties and thermal conductivity. It should be noted that the accuracy of MD estimations strongly depends on the accuracy of the interatomic potentials, so the thermal and mechanical properties estimations may vary, based on the choice of interatomic potentials. For example, the experimentally measured thermal conductivity of graphene is in the range 1500-5300 W/mK [12–14], while the thermal conductivity calculated by MD simulation predicted 350-3000 W/mK for this nanomaterial [15,16]. On the other hand, when using DFT-BTE methods, properties of materials may change depending on the choice of exchange-correlation functional and computational details for calculating 2nd and 3rd order force-constants[17,18].

In recent years, significant advances in machine-learning techniques have created new opportunities to address challenges in a variety of fields, including materials science[19–21]. Machine-learning interatomic potentials have been successfully employed to predict novel materials [22] and to study lattice dynamics[23,24] and thermal conductivity of materials[25–28]. MD simulations based on interatomic potentials obtained from machine learning have been shown to provide DFT level accuracy for calculated energy and force levels. However, the disadvantage of the current method is that the training process requires several hundred single-point DFT calculations during the active learning process [23,25].

In this work, thermal conductivity and mechanical properties of $C_2N$ nanostructure are studied based on the passively trained machine-learning interatomic potentials. This process requires primary ab-initio molecular dynamics trajectories, without the need for DFT calculations. First, interatomic potentials are trained for $C_2N$ nanostructure. Then, the MLIP package is used to estimate the thermal conductivity and mechanical properties of $C_2N$ nanomaterial using non-equilibrium molecular dynamics simulations. The thermo-mechanical results are also compared to that of classical interatomic potential. Besides, the stability and mechanical properties of $C_2N$ in the presence of point defects are investigated using machine learning interatomic potential.

## 2. Methodology

In this work, we train MLIP interatomic potentials using Ab-initio molecular dynamics (AIMD) trajectories. Then, using classical MD simulations, the developed potentials are used to predict the thermal conductivity and mechanical properties of the $C_2N$ monolayer.

**Ab-initio modeling**

The Ab-initio calculations are performed to create the required training sets for the development of MLIPs. The AIMD simulations are performed for pristine phases of $C_2N$ in five different temperatures between 50K-1000K. Required training sets for mechanical properties are also obtained by AIMD simulations at eight different strain levels. The strain-induced structures are first optimized.

To provide training sets required for the point defect structure, the $C_2N$ is first optimized in all possible cases with a point defect, and then each optimized structure is simulated using AIMD. All AIMD simulations are done for 1000 timesteps.

## Training of interatomic potentials

Momentum tensor potential (MTP) is utilized to describe the atomic interactions [29]. MTPs are based on the representation of inertia tensors of atomic environments of various ranks multiplied by radial polynomial functions, which allows for many-body interactions. MTP parameters are trained in such a way that the difference between the predicted and the first principles energy, forces, and stresses reach a minimum value [29]:

$$\sum_{k=1}^{K}\left[w_e(E_k^{AIMD} - E_k^{MTP})^2 + w_f \sum_{i}^{N} |f_{k,i}^{AIMD} - f_{k,i}^{MTP}|^2 + w_s \sum_{i,j}^{N} |\sigma_{k,ij}^{AIMD} - \sigma_{k,ij}^{MTP}|^2\right] \to min \quad (1)$$

where $E_k^{AIMD}$, $f_{k,i}^{AIMD}$, and $\sigma_{k,ij}^{AIMD}$ are energy, atomic forces, and stresses in the training set, respectively, and $E_k^{MTP}$, $f_{k,i}^{MTP}$, and $\sigma_{k,ij}^{MTP}$ are the corresponding values calculated with MTP, K is the number of training set configurations, N is the number of atoms, and $w_e$, $w_f$, and $w_s$ are non-negative weights that denote the importance of energies, forces, and stresses, which are set to 1, 0.1, and 0.001, respectively. The training method is a nonlinear optimization that is solved using the Broyden-Fletcher-Goldfarb-Shanno (BFGS) algorithm [30].

Since the trajectories obtained by AIMD are correlated within a short time, 10% of the original trajectories are considered as subsample sets. Then, moment tensor potentials (MTPs)[29] are trained to describe the atomic interactions. In this work, one MTP is developed for both thermal and mechanical simulations. In the first step, MTPs are trained over subsample sets. Next, to evaluate the accuracy of the trained potentials, it is checked across all AIMD trajectories, and configurations with a high degree of extrapolation[31] are then added to the subsample sets. The final MTP is obtained by training over the new subsample sets. This MTP is then used to study the thermal and mechanical properties of pristine and defective $C_2N$ (Figure 1).

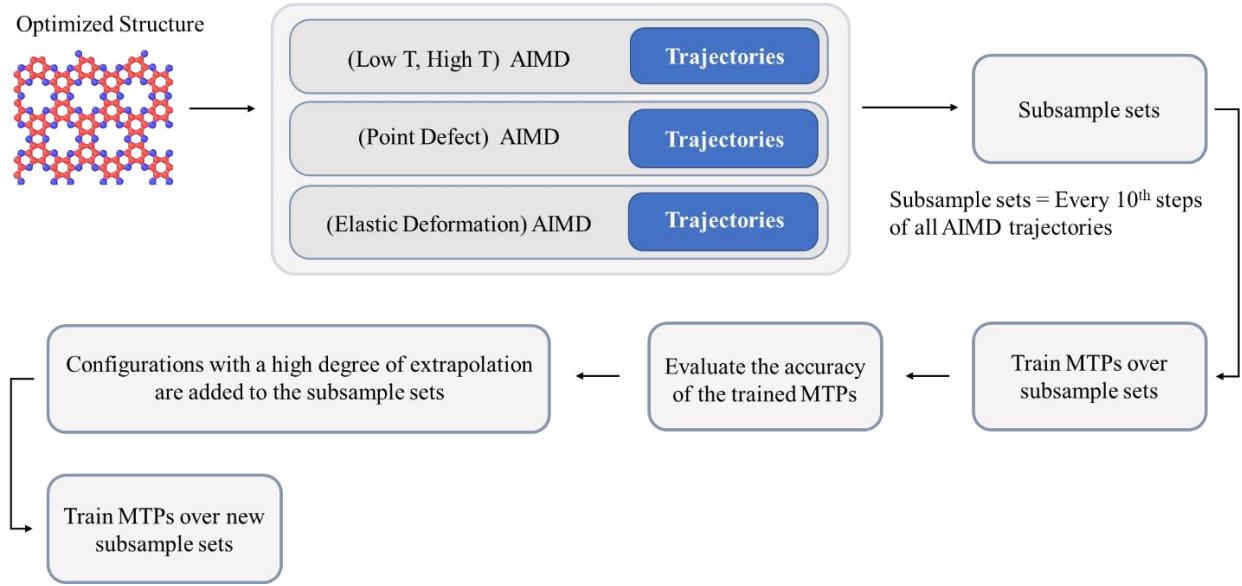

**Figure 1.** The algorithm of training moment tensor potential (MTP) using ab-initio molecular dynamics (AIMD) trajectories.

**Molecular dynamics modeling**

Non-equilibrium molecular dynamics (NEMD) simulations are performed to predict the lattice thermal conductivity and mechanical properties of the $C_2N$ monolayer, which is a reliable approaches to address thermal transport problems [32–35]. A large-scale Atomic/Molecular Massively Parallel Simulator (LAMMPS) package[36] is employed with MTP potential to introduce atomic interactions. To eliminate the effects of atoms at the boundary edges, $C_2N$ nanostructure with periodic boundary conditions is considered. Moreover, periodic boundary conditions minimize finite-length effects. All simulations are performed with an average temperature of 300K and a time increment of 0.5 fs.

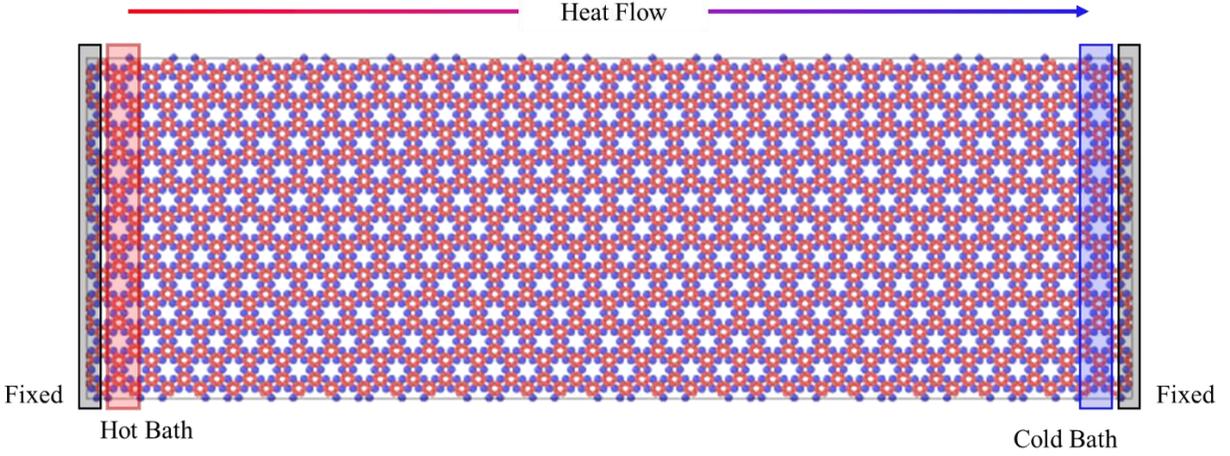

**Figure 2.** System setup for NEMD thermal conductivity calculations. A temperature gradient is imposed on the system by the hot and cold ends of the sample.

To calculate lattice thermal conductivity, the C$_2$N nanostructure was relaxed at room temperature for 2.5 ps using a Nose-Hoover barostat and thermostat (NPT). After obtaining the equilibrated structure, a temperature gradient is created by hot and cold areas on either side of the specimen which imposes heat flux through the nanostructure, as shown in Figure 2. A microcanonical ensemble (NVE) for 3ns is used to calculate the average energy added and subtracted from the two baths.

Figure 3(a) represents the accumulative energy variations of hot and cold domain, respectively, as a function of time. The net amount of energy exchanged between hot and cold region must be zero over time, and therefore the system reaches a stationary state. Figure 3(b) shows the temperature profile within the structure. By ignoring nonlinear temperature jumps adjacent to the hot and cold regions, a linear temperature gradient is established along the sample. Finally, utilizing Fourier's heat conduction equation, the thermal conductivity is calculated:

$$\kappa = -q'' / \frac{dT}{dx} \qquad (2)$$

where $q''$ is the heat flux between the cold and hot regions and $\frac{\partial l}{\partial T}$ denotes the temperature gradient within the system.

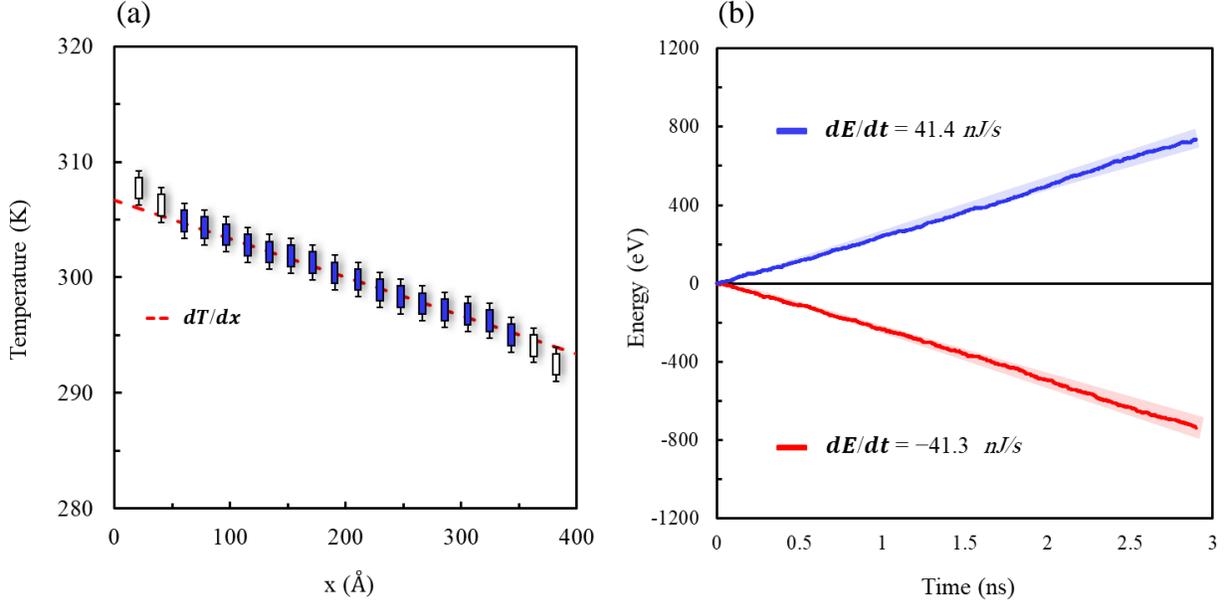

**Figure 3.** (a) Accumulative energy variations of the hot and cold domain (b) Temperature profile in the system after transient period.

To calculate Young's modulus and shear modulus, the length of the simulation box is increased along the loading direction at a constant engineering strain rate of $1\times10^9$ s$^{-1}$. The stress components (S) are calculated by the virial theorem[37]:

$$S = \frac{1}{V}\sum_{a\in V}\left[-m\vec{v}_a \otimes \vec{v}_a + \frac{1}{2}\sum_{a\neq b}(\vec{r}_{ab} \otimes \vec{F}_{ab})\right] \qquad (3)$$

where $m$ and $\vec{v}_a$ represent the mass and velocity of the atom $a$, respectively, $\vec{r}_{ab}$ is the position vector of the atom $a$ with respect to atom $b$, $\vec{F}_{ab}$ represents the force vector on atom $a$ due to $b$, $V$ represents the volume of the C2N nanostructure and $\otimes$ represents the product of the tensor.

## 3. Results and Discussion

The focus of this work is on the development of MLIP for modeling the thermal conductivity and mechanical properties of $C_2N$ monolayer using the classical MD. For this purpose, the training set must include all possible configurations that may occur during classical MD simulations. Accordingly, to create training data sets, AIMD simulations are performed at different temperatures, from 50 K to 1000 K, with a total simulation time of 4 ps. Low-temperature simulations can be useful for capturing low-displacement modes equivalent to long-wavelength phonons, while higher-temperature simulations can be useful for describing high-frequency optical modes. Moreover, since NEMD simulations involve relatively long-length $C_2N$ sheets, AIMD paths at higher temperatures are required to involve large local deformations that may occur during long-length computations.

In Figure 4, phonon dispersion relation (PDR) predicted by MTP and DFT are compared. As it can be seen, significant agreement is reached between MTP and DFT results. While the acoustic modes are very well reproduced by MTP, slight deviations are noticed for the optical modes in the $C_2N$ monolayer. To better reproduction of optical modes, the training set should be improved by considering AIMD trajectories at more range of temperatures, particularly at higher temperatures.

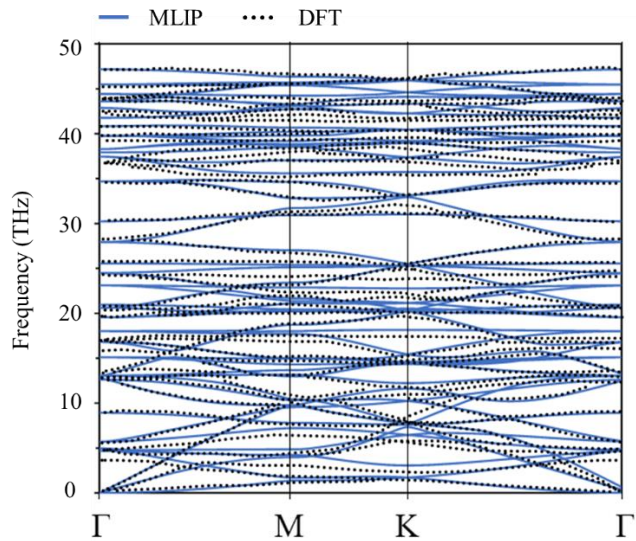

**Figure 4.** Phonon dispersion relations of $C_2N$ are acquired by MTP (continuous blue lines) and DFT [27] (black-dotted lines).

**Thermal Conductivity**

In this subsection, the lattice thermal conductivity of $C_2N$ is investigated by NEMD based on a passively trained MLIP and compared with those obtained with classical interatomic potentials (Tersoff optimized by Lindsay [38]). Figure 5 illustrates the effect of length on the thermal conductivity of the $C_2N$ monolayer at room temperature. As a characteristic of two-dimensional materials, thermal conductivity at small lengths is highly dependent on length due to ballistic thermal transport. Normally, the thermal conductivity of 2D nanomaterials eventually converges to reach the diffuse heat transfer regime at higher lengths. Bulk thermal conductivity of 2D materials, as well as its effective phonon MFP, can be estimated, using the following equation [39]:

$$\frac{1}{k_l} = \frac{1}{k_\infty}\left(1 + \frac{\lambda}{L}\right) \qquad (4)$$

where $k_\infty$ is the thermal conductivity of the material at infinite length, and $\lambda$ is an effective phonon mean free path (MFP) for the material, and L is the finite length of the sample.

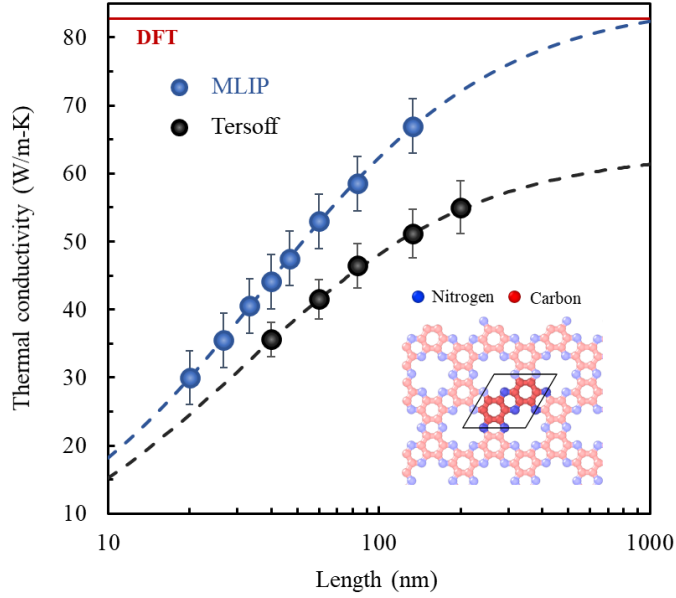

**Figure 5.** The lattice thermal conductivity of $C_2N$ as a function of the length at 300 K: comparison of MLIP, Tersoff potential and DFT-BTE calculation [40].

The predicted lattice thermal conductivity, at room temperature, is 85.47 ± 3 W/m-K by MTPs and 63.29 ± 5 W/m-K by Tersoff. The results suggest that the thermal conductivity, obtained by passively trained MTPs, shows a better agreement compared to the DFT-BTE result [40]. The values of $k_\infty$ and $\lambda$ are also summarized in Table 1.

Figure 6 shows the thermal conductivity ($k_\infty$) obtained from eq. 4, as a function of temperature between 300 K and 800 K. In this temperature range, the thermal conductivity of a bulk crystalline material is expected to be inversely proportional to temperature [41]. In other words, when the only source of thermal resistance is phonon-phonon scattering then $\kappa \sim T^{-1}$. As it can be seen from figure 6, the MLIP results has a better match to $\kappa \sim T^{-1}$ fitting.

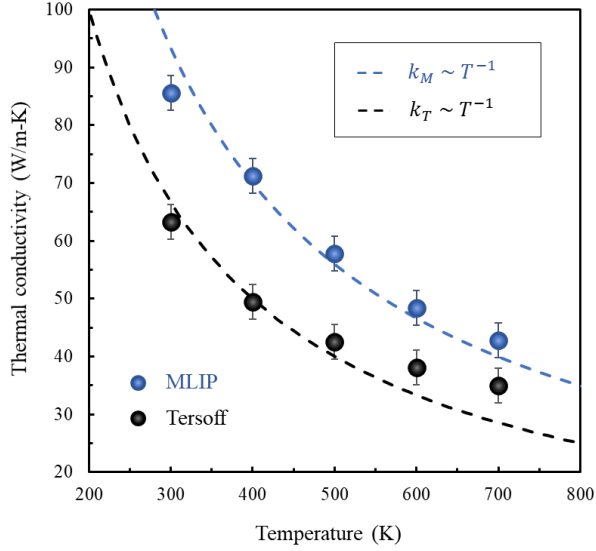

**Figure 6.** Temperature-dependent lattice thermal conductivity of single-layer $C_2N$: comparison of MLIP and Tersoff potentials.

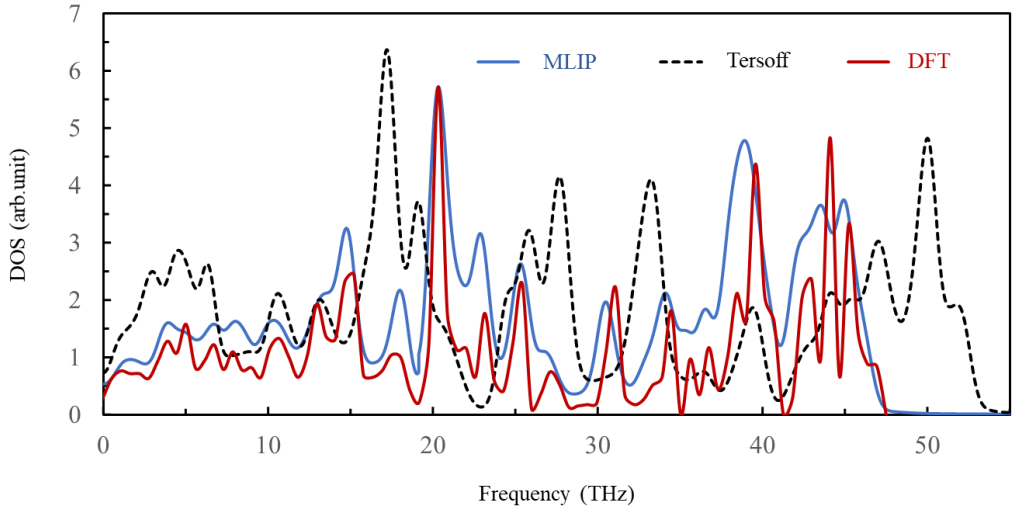

**Figure 7.** PDOS of the single-layer C2N: comparison of DFT, MLIP and Tersoff potentials.

Figure 7 shows the phonon density of states (PDOS) results obtained by three methods of DFT, MD with MLIP and MD with Tersoff potential calculated at T=1K. The DFT and MLIP results are more consistent than the classical Tersoff potential results. The PDOS were calculated using Fourier's transform of atomic velocity autocorrelation function. As it turns out, the Tersoff potential function shows frequencies that do not exist in the DFT results. However, these frequencies have high values and are related to the optical phonon branch and do not contribute much to the thermal conductivity of the structure.

**Mechanical properties**

We performed MD simulations of uniaxial tension at 300K. The NEMD results for the stress-strain response of $C_2N$ under uniaxial loading are shown in Figure 8. In this case, we found an elastic modulus of around 390 ± 3 GPa for machine learning interatomic potential and 359 ± 5 GPa for classic interatomic potential, which corresponds to a 5% and 14% difference from the prediction based on DFT (410.5 ± 5 GPa [42]), respectively. It should be noted that the Tersoff potential has been developed for heat conduction in graphene and hexagonal boron-nitride layers [31,32] and has not been developed to exactly reproduce the elastic constants [33]. On the other hand, since the MLIP has been practiced at different strains, the results show better agreement with DFT results compared to classical Tersoff potential. This way, at room temperature, the ultimate strength and fractural strain are found to be around 42 ± 4 GPa and 0.29 ± 0.1, respectively.

**Table 1.** Comparison of Intrinsic thermal conductivity, mean free path, elastic modulus, ultimate strength, and fractural strain for $C_2N$ that obtained from MD with MTP and Tersoff potentials and DFT-BTE method

| Method | MD (MTP) | MD (Tersoff) | DFT |
|---|---|---|---|
| Thermal Conductivity (W/m-K) | 85.47 ± 3 | 63.29 ± 5 | 82.22 [40] |
| Mean Free Path (nm) | 36.73 ± 1 | 31.54 ± 1 | 36 [40] |
| Elastic Modulus (GPa) | 390 ± 3 | 359 ± 5 | 410.5 ± 5 [42] |
| Ultimate Strength (GPa) | 42 ± 4 | 58.5 ± 7 | - |
| Fractural Strain | 0.29 ± 0.1 | 0.21 ± 0.1 | - |

Figure 8(b) shows the local stress of the C$_2$N nanostructure under uniaxial tension. In contrast to the results obtained from the classical Tersoff potential, the MTP shows a gradual rupture in the stress-strain curve. The fracture process ultimately involves the rupture of atomic bonds, which is shown in figure 8(b).

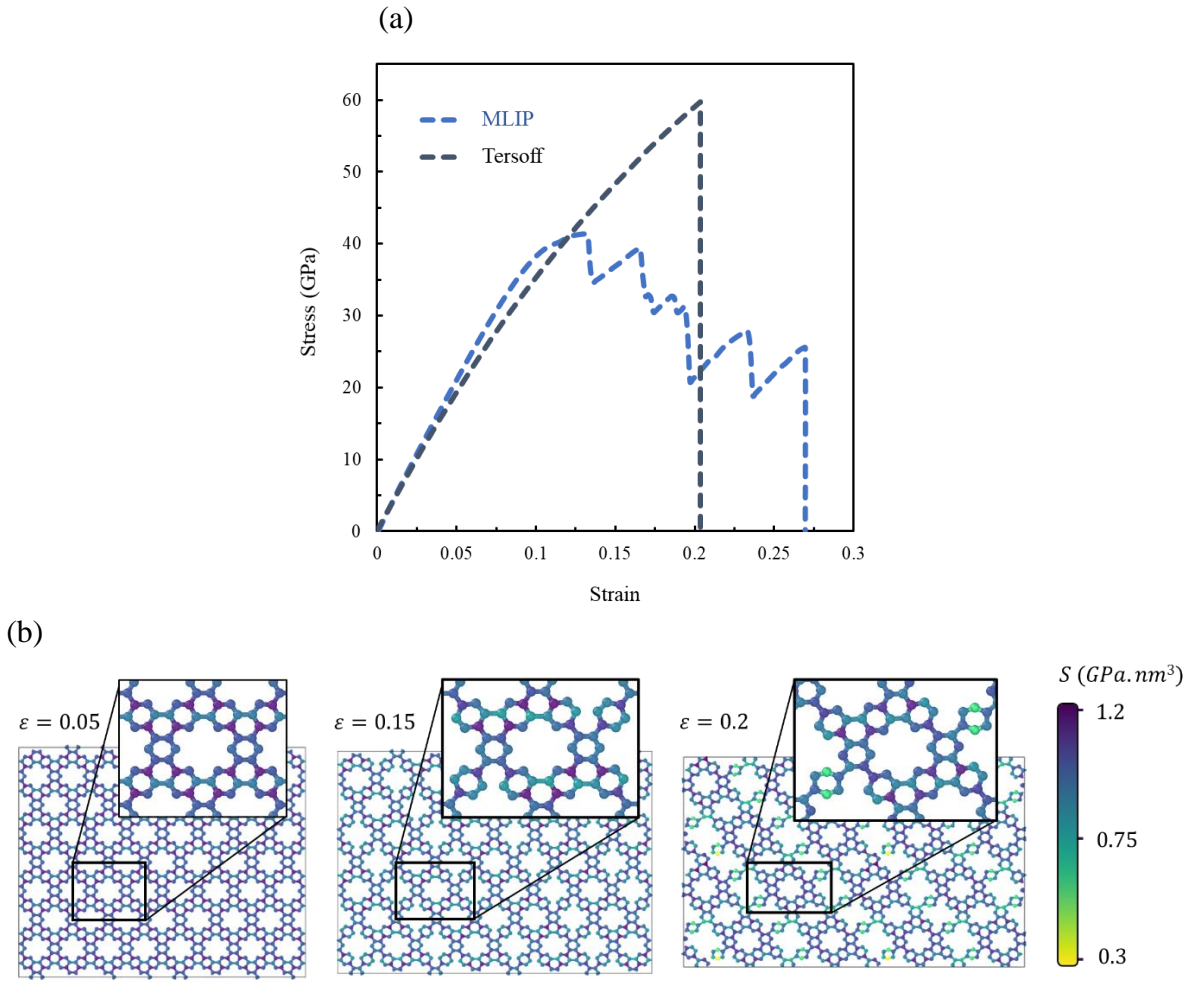

**Figure 8. a)** Stress-strain relation of pristine C$_2$N under uniaxial tensile load: comparison of MLIP and Tersoff potentials. **b)** Uniaxial deformation process of pristine C$_2$N at various engineering strain levels along zigzag direction obtained using MLIP. The color bar shows the stress values multiplied by the volume of the system.

To show the capability of MLIP to study the mechanical properties, the effect of point defects on mechanical properties of the C$_2$N nanostructure is also investigated. Figure 9(a) shows the stress-strain curve in both pristine and defective forms. As shown in Figure 9(a), the mechanical properties are reduced in the presence defects. With 2% defects, the elastic modulus, fractural stress, and ultimate strength are reduced by 5%, 40%, and 15%, respectively. Figure 9(b) also shows that the structure with point defects fails around the defect where there is a stress concentration and broken atomic bonds. It reveals that the MLIP can properly predict the mechanical properties of both pristine and defective C2N sheets.

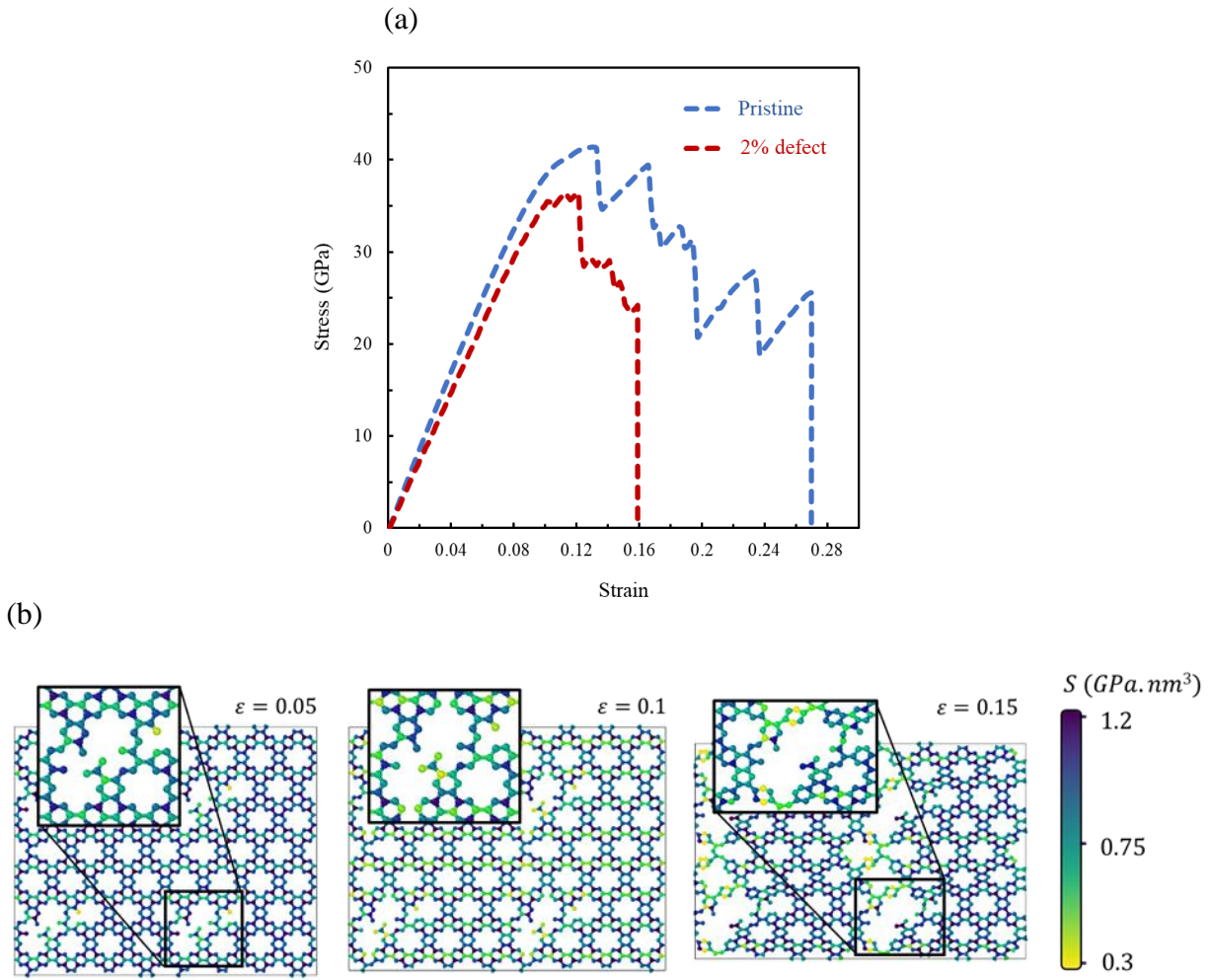

**Figure 9. a)** Stress-strain relation of defective $C_2N$ under uniaxial tensile load. **b)** Uniaxial deformation process of single-layer defective $C_2N$ at various engineering strain levels along zigzag direction. The color bar shows the stress values multiplied by the volume of the system.

4. **Conclusion**

Resorting to non-equilibrium molecular dynamics (NEMD), machine learning interatomic potential (MLIP) is introduced as a promising potential function for predicting the thermal and mechanical properties of monolayer $C_2N$. Concluding remarks can be summarized as follows:

- Passively trained MLIP over short ab-initio trajectories is able to well regenerate the phononic properties in close agreement with those by DFT-based results.
- It is found that the thermal conductivity of monolayer $C_2N$ is considerably length-dependent. The thermal conductivity of balk $C_2N$, at room temperature, is predicted as $85.47 \pm 3$ W/m-K by MLIP and $63.29 \pm 5$ W/m-K by Tersoff potential. The thermal conductivity optioned using MLIP predicted a closer value to that of DFT (82.22 W/m-K).
- Passively trained MLIP over various elastic deformation ab-initio trajectories is able to predict mechanical properties of $C_2N$ monolayer. Using MLIP the elastic modulus, ultimate strength, and fractural strain are estimated to be $390 \pm 3$ GPa, $42 \pm 4$ GPa, and $0.29 \pm 0.1$, respectively.
- While the stability of defective $C_2N$ using Tersoff potential is in doubt; the MLIP is well predicting the properties of $C_2N$ with point defects. Applying 2% defects, the elastic modulus, fractural stress, and ultimate strength are reduced by 5%, 40%, and 15%, respectively.
- While the MLIP offers accurate prediction, close to DFT, in both mechanical and thermal properties; the classical interatomic potentials, e.g. Tersoff, are faster ways to predict 2D nanomaterials properties.

The results of the current study shed light on the unparalleled potential of machine learning interatomic potentials for estimating the thermo-mechanical properties of two-dimensional nanostructures, including $C_2N$.

**References**

[1] G.R. Bhimanapati, Z. Lin, V. Meunier, Y. Jung, J. Cha, S. Das, D. Xiao, Y. Son, M.S. Strano, V.R. Cooper, others, Recent advances in two-dimensional materials beyond graphene, ACS Nano. 9 (2015) 11509–11539.

[2] K.S. Novoselov, A.K. Geim, S. V Morozov, D. Jiang, Y. Zhang, S. V Dubonos, I. V Grigorieva, A.A. Firsov, Electric field effect in atomically thin carbon films, Science (80-. ). 306 (2004) 666–669.

[3] A.J. Mannix, X.-F. Zhou, B. Kiraly, J.D. Wood, D. Alducin, B.D. Myers, X. Liu, B.L. Fisher, U. Santiago, J.R. Guest, M.J. Yacaman, A. Ponce, A.R. Oganov, M.C. Hersam, N.P. Guisinger, Synthesis of borophenes: Anisotropic, two-dimensional boron polymorphs., Science. 350 (2015) 1513–6. doi:10.1126/science.aad1080.

[4] L. Li, Y. Yu, G.J. Ye, Q. Ge, X. Ou, H. Wu, D. Feng, X.H. Chen, Y. Zhang, Black phosphorus field-effect transistors, Nat. Nanotechnol. 9 (2014) 372.

[5] S. Arabha, A.H. Akbarzadeh, A. Rajabpour, Engineered porous borophene with tunable anisotropic properties, Compos. Part B Eng. 200 (2020) 108260.

[6] H. Ghasemi, A. Rajabpour, A.H. Akbarzadeh, Tuning thermal conductivity of porous graphene by pore topology engineering: Comparison of non-equilibrium molecular dynamics and finite element study, Int. J. Heat Mass Transf. 123 (2018) 261–271.

[7] S. Arabha, A. Rajabpour, Effect of planar torsional deformation on the thermal conductivity of 2D nanomaterials: A molecular dynamics study, Mater. Today Commun. 22 (2020) 100706.

[8] J. Mahmood, E.K. Lee, M. Jung, D. Shin, I.-Y. Jeon, S.-M. Jung, H.-J. Choi, J.-M. Seo, S.-Y. Bae, S.-D. Sohn, others, Nitrogenated holey two-dimensional structures, Nat. Commun. 6 (2015) 1–7.

[9] L. Zhu, Q. Xue, X. Li, T. Wu, Y. Jin, W. Xing, C 2 N: an excellent two-dimensional monolayer membrane for He separation, J. Mater. Chem. A. 3 (2015) 21351–21356.

[10] W. Cao, H. Xiao, T. Ouyang, J. Zhong, Ballistic thermoelectric properties of nitrogenated holey graphene nanostructures, J. Appl. Phys. 122 (2017) 174302.

[11] R.M. Tromer, A. Freitas, I.M. Felix, B. Mortazavi, L. Machado, S. Azevedo, L.F.C. Pereira, Electronic, optical and thermoelectric properties of boron-doped Nitrogenated Holey Graphene, Phys. Chem. Chem. Phys. (2020).

[12] D.S. Ghosh, I. Calizo, D. Teweldebrhan, E.P. Pokatilov, D.L. Nika, A.A. Balandin, W. Bao, F. Miao, C.N. Lau, Extremely high thermal conductivity of graphene: Prospects for thermal management applications in nanoelectronic circuits, Appl. Phys. Lett. 92 (2008) 151911.

[13] A.A. Balandin, S. Ghosh, W. Bao, I. Calizo, D. Teweldebrhan, F. Miao, C.N. Lau, Superior Thermal Conductivity of Single-Layer Graphene, Nano Lett. 8 (2008) 902–907. doi:10.1021/nl0731872.

[14] L.A. Jauregui, Y. Yue, A.N. Sidorov, J. Hu, Q. Yu, G. Lopez, R. Jalilian, D.K. Benjamin, D.A. Delkd, W. Wu, others, Thermal transport in graphene nanostructures: Experiments and simulations, Ecs Trans. 28 (2010) 73.

[15] J.A. Thomas, R.M. Iutzi, A.J.H. McGaughey, Thermal conductivity and phonon transport in empty and water-filled carbon nanotubes, Phys. Rev. B. 81 (2010) 45413.

[16] B. Mortazavi, T. Rabczuk, Multiscale modeling of heat conduction in graphene laminates, Carbon N. Y. 85 (2015) 1–7.

[17] L. Lindsay, D.A. Broido, N. Mingo, Flexural phonons and thermal transport in graphene, Phys. Rev. B. 82 (2010) 115427.

[18] S. Huang, M. Segovia, X. Yang, Y.R. Koh, Y. Wang, D.Y. Peide, W. Wu, A. Shakouri, X. Ruan, X. Xu, Anisotropic thermal conductivity in 2D tellurium, 2D Mater. 7 (2019) 15008.

[19] R. Magar, P. Yadav, A.B. Farimani, Potential neutralizing antibodies discovered for novel corona virus using machine learning, ArXiv Prepr. ArXiv2003.08447. (2020).

[20] J. Schmidt, M.R.G. Marques, S. Botti, M.A.L. Marques, Recent advances and applications of machine learning in solid-state materials science, Npj Comput. Mater. 5 (2019) 1–36.

[21] G.R. Schleder, A.C.M. Padilha, C.M. Acosta, M. Costa, A. Fazzio, From DFT to machine learning: recent approaches to materials science--a review, J. Phys. Mater. 2 (2019) 32001.

[22] E. V Podryabinkin, E. V Tikhonov, A. V Shapeev, A.R. Oganov, Accelerating crystal structure prediction by machine-learning interatomic potentials with active learning, Phys. Rev. B. 99 (2019) 64114.

[23] V. V Ladygin, P.Y. Korotaev, A. V Yanilkin, A. V Shapeev, Lattice dynamics simulation using machine learning interatomic potentials, Comput. Mater. Sci. 172 (2020) 109333.

[24] M. Jafary-Zadeh, K.H. Khoo, R. Laskowski, P.S. Branicio, A. V Shapeev, Applying a machine learning interatomic potential to unravel the effects of local lattice distortion on the elastic properties of multi-principal element alloys, J. Alloys Compd. 803 (2019) 1054–1062.

[25] P. Korotaev, I. Novoselov, A. Yanilkin, A. Shapeev, Accessing thermal conductivity of complex compounds by machine learning interatomic potentials, Phys. Rev. B. 100 (2019) 144308.

[26] B. Mortazavi, E. V Podryabinkin, S. Roche, T. Rabczuk, X. Zhuang, A. V Shapeev, Machine-learning interatomic potentials enable first-principles multiscale modeling of lattice thermal conductivity in graphene/borophene heterostructures, Mater. Horizons. 7 (2020) 2359–2367.

[27] B. Mortazavi, I.S. Novikov, E. V Podryabinkin, S. Roche, T. Rabczuk, A. V Shapeev, X. Zhuang, Exploring phononic properties of two-dimensional materials using machine learning interatomic potentials, Appl. Mater. Today. 20 (2020) 100685.

[28] B. Mortazavi, F. Shojaei, M. Shahrokhi, M. Azizi, T. Rabczuk, A. V Shapeev, X. Zhuang, Nanoporous $C_3N_4$, $C_3N_5$ and $C_3N_6$ nanosheets; novel strong semiconductors with low thermal conductivities and appealing optical/electronic properties, Carbon N. Y. 167 (2020) 40–50.

[29] A. V Shapeev, Moment tensor potentials: A class of systematically improvable interatomic potentials, Multiscale Model. Simul. 14 (2016) 1153–1173.

[30] J.D. Head, M.C. Zerner, A Broyden—Fletcher—Goldfarb—Shanno optimization procedure for molecular geometries, Chem. Phys. Lett. 122 (1985) 264–270.

[31] E. V Podryabinkin, A. V Shapeev, Active learning of linearly parametrized interatomic potentials, Comput. Mater. Sci. 140 (2017) 171–180.

[32] Y. Hu, T. Feng, X. Gu, Z. Fan, X. Wang, M. Lundstrom, S.S. Shrestha, H. Bao, Unification of nonequilibrium molecular dynamics and the mode-resolved phonon Boltzmann equation for thermal transport simulations, Phys. Rev. B. 101 (2020) 155308.

[33] F. Müller-Plathe, A simple nonequilibrium molecular dynamics method for calculating the thermal conductivity, J. Chem. Phys. 106 (1997) 6082–6085.

[34] A. Rajabpour, S.M. Vaez Allaei, F. Kowsary, Interface thermal resistance and thermal rectification in hybrid graphene-graphane nanoribbons: A nonequilibrium molecular dynamics study, Appl. Phys. Lett. 99 (2011) 2011–2014. doi:10.1063/1.3622480.

[35] B. Mortazavi, S. Ahzi, Thermal conductivity and tensile response of defective graphene: A molecular dynamics study, Carbon N. Y. 63 (2013) 460–470. doi:10.1016/j.carbon.2013.07.017.

[36] S. Plimpton, Fast parallel algorithms for short-range molecular dynamics, J. Comput. Phys. 117 (1995) 1–19. doi:10.1006/jcph.1995.1039.

[37] J.A. Zimmerman, E.B. Webb, J.J. Hoyt, R.E. Jones, P.A. Klein, D.J. Bammann, Calculation of stress in atomistic simulation, in: Model. Simul. Mater. Sci. Eng., 2004. doi:10.1088/0965-0393/12/4/S03.

[38] L. Lindsay, D.A. Broido, Optimized Tersoff and Brenner empirical potential parameters for lattice dynamics and phonon thermal transport in carbon nanotubes and graphene, Phys. Rev. B. 81 (2010) 205441. doi:10.1103/PhysRevB.81.205441.

[39] P.K. Schelling, S.R. Phillpot, P. Keblinski, Comparison of atomic-level simulation methods for computing thermal conductivity, Phys. Rev. B. 65 (2002) 144306.

[40] T. Ouyang, H. Xiao, C. Tang, X. Zhang, M. Hu, J. Zhong, First-principles study of thermal transport in nitrogenated holey graphene, Nanotechnology. 28 (2016) 45709.

[41] J.M. Ziman, Electrons and phonons: the theory of transport phenomena in solids, Oxford university press, 2001.

[42] B. Mortazavi, O. Rahaman, T. Rabczuk, L.F.C. Pereira, Thermal conductivity and mechanical properties of nitrogenated holey graphene, Carbon N. Y. 106 (2016) 1–8.